\documentclass[12pt,thmsa]{article}

\usepackage{amsfonts}
\usepackage{graphicx}
\usepackage{epsfig}
\usepackage{graphics}

\makeatletter
\newcommand{\row}[1]%
{\mathord{\buildrel{\lower3pt%
\hbox{$\scriptscriptstyle\rightarrow$}}\over #1}}

\newcommand{\dyadic}[1]{\mathord{\dyadic@rrow{#1}}}
\newcommand{\dyadic@rrow}[1]{
\begin{picture}(12,12)(-1,0)
\put(-1,9){\makebox(0,0)[t]{$\scriptscriptstyle\downarrow$}}
\put(-1,9){\makebox(0,0)[l]{$\scriptscriptstyle\longrightarrow$}}
\put(5,0){\makebox(0,0)[b]{$#1$}}
\end{picture}
}

\newcommand{\bra}[1]{\bigl\langle #1 \bigr|}
\newcommand{\ket}[1]{\bigl| #1 \bigr\rangle}

\newcommand{\tr}[2][\,]{\mathrm{tr}_{#1}\left\{#2\right\}}

\begin{document}

\begin{center}
{\Large Immunity, Improving  and Retrieving the lost entanglement
of accelerated qubit-qutrit system  via local Filtering}\\
 {N. Metwally\\}
$^1$ Department of Mathematics, College of Science, Bahrain
University, Bahrain \\
$^2$Department of Mathematics, Faculty of Science,
Aswan University, Aswan, Egypt \\
email:nmetwally@gmail.com,
% nmetwally@uob.edu.bh \\
%Tel:00973 33237474, ~ Fax: 00973 17449145
\end{center}
\date{\today }

\begin{abstract}
 The possibility of immunized and improving the  entanglement  of
accelerated systems via local filtering is discussed. The maximum
bounds of entanglement depend on the dimensions of the accelerated
and the filtered subsystems. If the small dimensional subsystem is
accelerated and the large dimensional subsystem is filtered, one
can get a long-lived entanglement. Moreover, if the larger
subsystem is accelerated, then by filtering any subsystem, the
upper bounds of entanglement of the filtered state are larger then
that for the initial states.For any   accelerated subsystem, the
entanglement always increases as the filter strength of the large
dimensional subsystem  increases.

\end{abstract}

\topmargin=0.001cm \textheight=24cm \textwidth=17cm
\section{Introduction}
It  is well known that, to perform quantum information tasks,
maximum entangled states are required. Although, it is possible to
generate such  states, mainting  their isolation is a very
difficult task. There are many studies  devoted to investigate the
behavior of entanglement in different environments \cite{Eberly}.
On the other hand, the possibility of using these noise states to
perform some quantum information is studied from different point
of views\cite{Metwally09}.

Due to the interaction with the environments  the amount of
entanglement decreases and accordingly their efficiency to perform
quantum information tasks decreases. Therefore, it is  a necessity
to improve the weakness of entanglement. For this aim, there are
different protocols of purifications that have been  introduced
\cite{Bennt,Deutsch, Metwally2002, Metwally2006}, where the main
idea of these protocols is  based on local quantum operations,
classical communication and measurements. So, one can get a
smaller numbers  of strongly entangled  qubits from a large number
of weakly entangled ones. Bennett. et.al. \cite{Bennt1},  showed
that local filtering can be used to increase the entanglement of a
two-qubit
 by applying it on one of its subsystems.   A single local filter
 operation can be used to retrieve the loss of entanglement of
 particles passing through noise channels \cite{Siomau}. Moreover
 local filtering is used to increase the entanglement of subsystem
 at the expense of the other subsystem \cite{Sidha}. The effect of
 local filtering on the dynamics of some measures of quantum
 correlation of  a general two qubit state is discusses by
 Karmakar et. al, \cite{Karm}.

Some efforts have been done  recently to investigate the survival
amount of entanglement between different accelerated systems
\cite{Wang011}. These  accelerated states can be used to perform
some quantum information tasks such  as teleportation
\cite{Metwally2013} and quantum coding \cite{Sagheer}. Due to the
acceleration, the entanglement between the accelerated partners
decreases, where the decay rate depends on the initial
acceleration and the dimension of the accelerated subsystem.
Therefore, it is worth   to investigate, the possibility of
improving the entanglement between the accelerated particles  by
filtering one of the subsystems. We introduce this idea by using a
state composite of two different dimensional subsystems; qubit
$(2D)$ and qutrit $(3D)$. This state is  described by one
parameter, known as one parameter family\cite{Ann}, where it is
shown that, if the larger subsystem is accelerated the rate of
entanglement decay is larger than that depicted for accelerating
the small dimension subsystem \cite{metwally2016}.

This paper  is organized as follows: In Sec.2, we define the
qubit-qutrit  state and its  final  form in the non-inertial frame
when  one or both particles are  accelerated. The filtering
process is discussed in Sec.3, where  analytical solutions are
obtained for the final filtered states. The degree of entanglement
between the accelerated partner is quantified by using the
negativity as a measure of entanglement. Finally, we summarize or
results and conclusions in Sec. 4.

\section{ The suggested model}
 We consider a  state  that represents a class of qubit-qutrit system in $2\times
 3$ dimensional and  described by only one parameter.
 In the computation basis $00,01.02,10,11,12$, it takes the form

 \begin{eqnarray}
 \rho&=&\Bigl\{\frac{\mu}{2}(\ket{0}_b\bra{0}+\ket{1}_b\bra{1})\Bigl\}\otimes\ket{1}_t\bra{1}
 +\Bigl\{\frac{\mu}{2}\ket{1}_b\bra{1}+\frac{1-2\mu}{2}\ket{0}_b\bra{0}\Bigr\}\otimes\ket{2}_t\bra{2}
 \nonumber\\
 &&+\Bigl\{\frac{\mu}{2}\ket{0}_b\bra{1}+\frac{1-2\mu}{2}\ket{1}_b\bra{0}\Bigr\}\otimes\ket{0}_t\bra{2}
 +\Bigl\{\frac{\mu}{2}\ket{1}_b\bra{0}+\frac{1-2\mu}{2}\ket{0}_b\bra{1}\Bigr\}\otimes\ket{2}_t\bra{0}
 \nonumber\\
 &&+\Bigl\{\frac{\mu}{2}\ket{0}_b\bra{0}\Bigr\}\otimes\ket{0}_t\bra{0}+
 \Bigl\{\frac{1-2\mu}{2}\ket{1}_b\bra{0}\Bigr\}\otimes\ket{0}_t\bra{1}
\end{eqnarray}
where, $0\leq \mu\leq \frac{1}{2}$ and the subscript $"b"$ refers
to the qubit while $"t"$ refers to the qutrit
\cite{Guo,metwally2016}

In what follows, we consider  only one subsystem is accelerated.
Accordingly we have the following cases:accelerated  and filtered
qubit, accelerated qubit and filtered qutrit, accelerated qutrit
and filtered qubit and accelerated  and filtered qutrit.

\subsection{Accelerated the subsystems}
First, we  review the relation between Monkowski and Rindler
spaces \cite{Edu, Walls}. For {\it qubit systems}, if the
coordinates of a particle is defined by $(t,z)$ in Minkowski
space, then in Rindlier space it is defined by $(\tau,x)$, where
\begin{equation}
\tau=r \tanh(t/z),\quad x=\sqrt{t^2-z^2}, \quad -\infty<r<\infty,
\quad -\infty<x<\infty.
\end{equation}
Also, the annihilation operators $a_{ku}$ and $b_{-kU}$ in
Minkowski space can be written  by means of Rindler operators
\cite{un, Jason2013} $(c_{kR}^{(I)}, \quad d^{II}_{-kL})$ as,
\begin{eqnarray}\label{op}
a_{kU}&=&cosr_bc^{(I)}_{kR}-e^{-i\phi}\sin r_b d^{II}_{-KL},\quad
\nonumber\\
b^{\dagger}_{-kU}&=&e^{-i\phi}\sin r_bc^{(I)}_{kR}+\cos r_b
d^{II}_{-KL},
\end{eqnarray}
where $\tan r_b=Exp[-\pi\omega\frac{c}{a}]$, ~$0\leq r_b\leq
\pi/4$, $-\infty\leq a\leq\infty$, $\omega$ is the frequency,$c$
is the speed of  light, and $\phi$ is the phase space which can be
absorbed in the definition of the operators\cite{Jason2013}. The
relations $(3)$, mix a particle in the  region $I$ and its
anti-particle in the region $II$ such that the computational basis
$\ket{0_k}$and $\ket{1_k}$  can be written as\cite{metwally2016},
\begin{eqnarray}\label{Min}
\ket{0_k}&=&\cos r_b\ket{0_k}_I\ket{0_{-k}}_{II}+ \sin
r_b\ket{1_k}_I\ket{1_{-k}}_{II}, \nonumber\\
\ket{1_k}&=&a^\dagger_k\ket{0_k} =\ket{1_k}_I\ket{0_k}_{II}.
\end{eqnarray}
Now, for {\it qutrit systems}, the Minkowski vacuum state
$\ket{0_m}$, spin up state $\ket{\mathcal{U}}$, spin down state
$\ket{\mathcal{D}}$ and the pair state $\ket{\mathcal{P}}$,  in
the Rindler space are defined  in \cite{metwally2016,Alsing1,Juan}
as
\begin{eqnarray}
\ket{0_M}&=&\cos^2r_t\ket{0}_{I}\ket{0}_{II}+\frac{1}{2}e^{i\phi}\sin2
r_t(\ket{\mathcal{U}}_{I}\ket{\mathcal{D}}_{II}+\ket{\mathcal{D}}_{I}\ket{\mathcal{U}}_{II})
+e^{2i\phi}sin^2r_t\ket{\mathcal{D}}_{I}\ket{\mathcal{P}}_{II},
\nonumber\\
\ket{\mathcal{U}_M}&=&\cos
r_t\ket{\mathcal{U}}_I\ket{0}_{II}+e^{i\phi}\sin
r_t\ket{\mathcal{P}}_I\ket{\mathcal{U}}_{II},
\nonumber\\
\ket{\mathcal{D}_M}&=&\cos
r_t\ket{\mathcal{D}}_I\ket{0}_{II}-e^{i\phi}\sin
r_t\ket{\mathcal{P}}_I\ket{\mathcal{D}}_{II}.
\end{eqnarray}
By using the initial state (1) and the transformations (4), one
gets the final state of the accelerated system, where only the
qubit is accelerated. After tracing out the mode in region $II$,
the  final accelerated state between Alice and Bob in the first
region, $I$ can be written as,

\begin{eqnarray}
\rho^{ac-b}&=&\mathcal{A}_1\ket{00}\bra{00}+\mathcal{A}_2\ket{01}\bra{01}+
\mathcal{A}_3\ket{00}\bra{12}+\mathcal{A}_4\ket{12}\bra{00}
\nonumber\\
&&+\mathcal{A}_5\ket{10}\bra{02}+\mathcal{A}_6\ket{02}\bra{10}+\mathcal{A}_7\ket{02}\bra{02}+
\mathcal{A}_8\ket{10}\bra{10}
\nonumber\\
&&+\mathcal{A}_9\ket{12}\bra{12}+\ket{11}\bra{11}
\end{eqnarray}
where,
\begin{eqnarray}
\mathcal{A}_1&=&\mathcal{A}_2=\frac{\mu}{2}\cos^2r_b,\quad
\mathcal{A}_3=\mathcal{A}_4=\frac{\mu}{2}\cos r_b,
\nonumber\\
\mathcal{A}_5&=&\mathcal{A}_6=\frac{1-2\mu}{2}\cos r_b, \quad
\mathcal{A}_7=\frac{1-2\mu}{2}\cos^2 r_b,
\nonumber\\
\mathcal{A}_8&=&\frac{1-2\mu}{2}+\frac{\mu}{2}\sin^2 r_b,\quad
\mathcal{A}_9=\frac{\mu}{2}+\frac{1-2\mu}{2}\sin^2 r_b,\quad
\mathcal{A}_{10}=\frac{\mu}{2}(1+\sin^2 r_b).
\end{eqnarray}
Similarly, to  accelerate the qutrit one uses the transformation
(5) and  the initial state (1). However by tracing out the mode in
$II$, the final accelerated state between the partners in the
first region takes the form,
\begin{eqnarray}
\rho^{ac-t}&=&\mathcal{B}_{1}\ket{00}\bra{00}+\mathcal{B}_2\ket{01}\bra{01}+
\mathcal{B}_3\ket{02}\bra{02}+\mathcal{B}_4\ket{0\mathcal{P}}\bra{0\mathcal{P}}
\nonumber\\
&&+\mathcal{B}_{5}\ket{10}\bra{10}+\mathcal{B}_6\ket{11}\bra{11}+
\mathcal{B}_7\ket{12}\bra{12}+\mathcal{B}_8\ket{1\mathcal{P}}\bra{1\mathcal{P}}\nonumber\\
&&+\mathcal{B}_{9}\ket{12}\bra{00}+\mathcal{B}_{10}\ket{1\mathcal{P}}\bra{01}+
\mathcal{B}_{11}\ket{02}\bra{10}+\mathcal{B}_{12}\ket{0\mathcal{P}}\bra{11}
\nonumber\\
&&+\mathcal{B}_{13}\ket{00}\bra{12}+\mathcal{B}_{14}\ket{01}\bra{1\mathcal{P}}+
\mathcal{B}_{15}\ket{10}\bra{02}+\mathcal{B}_{16}\ket{11}\bra{0\mathcal{P}}\nonumber\\
\end{eqnarray}
where,
\begin{eqnarray}
\mathcal{B}_1&=&\frac{\mu}{2}\cos^4 r_t,\quad
\mathcal{B}_2=\frac{\mu}{2}\cos^2 r_t(1+\sin^2 r_t),\quad
\mathcal{B}_3=\frac{\mu}{8}\sin^22r_t,\quad
\nonumber\\
\mathcal{B}_4&=&\sin^2r_t\left(\frac{p}{2}\sin^2r_t+\frac{1-2\mu}{2}\right),\quad
\mathcal{B}_5=\frac{1-2\mu}{2}\cos^4r_t, \quad
\mathcal{B}_6=\cos^2r_t\left(\frac{\mu}{2}+\frac{1-2\mu}{2}\sin^2r_t\right),\quad
\nonumber\\
\mathcal{B}_7&=&\frac{1-2\mu}{8}\sin^22r_t,\quad
\mathcal{B}_8=\sin^2r_t\left(\frac{\mu}{2}+\frac{1-2\mu}{2}\sin^2r_t\right),\quad
\mathcal{B}_9=\frac{\mu}{2}\cos^3r_t,\quad \quad
\nonumber\\
\mathcal{B}_{10}&=&\frac{\mu}{4}\sin2r_t\sin r_t,\quad
\mathcal{B}_{11}=\frac{1-2\mu}{2}\cos^2r_t,\quad
\mathcal{B}_{12}=\frac{1-2\mu}{4}\sin2r_t\sin r_t,\quad\
\nonumber\\
\mathcal{B}_{13}&=&\mathcal{B}_9,\quad\mathcal{B}_{14}=\mathcal{B}_{10},\quad
\mathcal{B}_{15}=\mathcal{B}_{11},\quad\mathcal{B}_{16}=\mathcal{B}_{12}.
\end{eqnarray}

The entanglement is  quantified by  using  the negativity as a
measure, where for a system consists of two different dimensions
as qubit and qutrit, the negativity  of a bipartite system
consists of two subsystems have dimensions $d_1$ and
$d_2$,$(d_1<d_2)$ is given by

\begin{equation}
{\Large\mathcal{E}}=\frac{1}{d_1-1}\Bigl\{||\rho^{T_2}_{ab}||-1\Bigr\},
\end{equation}
where   $\rho^{T_b}_{ab}$  is the partial transpose with respect
to the largest dimension subsystem and $||.||$ is the trace
norm\cite{Ann,Karpat}.

\section{Filtering process}

In this section, we investigate the effect of the local filter
operation on the behavior of the degree of entanglement. For qubit
system the filter operation is defined by a non-trace preserving
operator \cite{Hofmann,Siomau}. In the computational basis the
filter operator $\mathcal{F}_b$ can be described by
\begin{equation}
\mathcal{F}_{b}=\sqrt{\kappa}\ket{0}\bra{0}+\sqrt{1-\kappa}\ket{1}\bra{1}
\end{equation}
where, $0<\kappa<1$. Let us assume that the accelerated state is
given by $\rho^{ac-q_i}$, where $i=b,t$. If the qubit is filtered
then the  filtered state is given by \cite{Siomau,Marco}
\begin{equation}
\rho_{b-F}^{ac-q_i}=\frac{1}{\mathcal{N}_b}\Bigl(\mathcal{F}_{b}\otimes
I_{3\times
3}\Bigr)\rho^{ac-q_i}\Bigl(\mathcal{F}^{\dagger}_{b}\otimes
I_{3\times 3}\Bigr)
\end{equation}
where $\mathcal{N}_b={\tr{ \Bigl(\mathcal{F}_{b}\otimes I_{3\times
3}\Bigr)\rho^{ac-q_i}\Bigl(\mathcal{F}^{\dagger}_{b}\otimes
I_{3\times 3}\Bigr)}}$ is a normalization factor. For qutrit
system the filter operator is defined by \cite{Yuma}
\begin{eqnarray}
\mathcal{F}_{t}^{(1)}&=&\ket{0}\bra{0}+\sqrt{1-\mathcal{Q}}\ket{1}\bra{1}+\sqrt{\mathcal{Q}}\ket{2}\bra{2},
\nonumber\\
\mathcal{F}_{t}^{(2)}&=&\sqrt{\mathcal{Q}}\ket{1}\bra{1}+\sqrt{1-\mathcal{Q}}\ket{2}\bra{2}.
\end{eqnarray}
If Bob applies the filter operation on his particle then the
output state is given by

\begin{equation}
\rho_{t-F}^{ac-q_i}=\frac{1}{\mathcal{N}_t}\sum_{j=1}^{2}\Bigl(I_{2\times
2}\otimes\mathcal{F}^{(j)}_{t}\Bigr)\rho^{ac-q_i}\Bigl(I_{2\times
2}\otimes\mathcal{F}^{\dagger(j)}_{t}\Bigr)
\end{equation}
where the normalization factor is  $\mathcal{N}_{t}={\tr{
\sum_{j=1}^{2}\Bigl(I_{2\times
2}\otimes\mathcal{F}^{(j)}_{t}\Bigr)\rho^{ac-q_i}\Bigl(I_{2\times
2}\otimes\mathcal{F}^{\dagger(j)}_{t}\Bigr)}}$, $j=1,2$.

\subsection{Alice' s qubit is accelerated}
\begin{itemize}
\item{\it Alice's qubit is  filtered\\} For this case, it is
assumed that only Alice filters her qubit by applying the operator
(11). The final filtered state is given by,
\begin{eqnarray}
\rho^{ac-b}_{b-F}&=&\frac{1}{\mathcal{N}_{b-F}}\Bigl\{\kappa\left(A_1\ket{00}\bra{00}+A_2\ket{01}\bra{01}+A_7\ket{02}\bra{02}\right)
\nonumber\\
&&+(1-\kappa)(A_8\ket{10}\bra{10}+A_9\ket{12}\bra{12}+
  A_{10}\ket{11}\bra{11})
\nonumber\\
&&+\sqrt{\kappa}\sqrt{1-\kappa}\left(A_3\ket{00}\bra{12}+A_4\ket{12}\bra{00}
+A_{5}\ket{10}\bra{02}+A_6\ket{02}\bra{10}\right)\Bigr\}
\end{eqnarray}
where, the coefficients $\mathcal{A}_i, i=1..10$ are given by (7)
and the normalized factor is
\begin{eqnarray*}
\mathcal{N}_{b-F}=\kappa\left(\mathcal{A}_1+\mathcal{A}_2+\mathcal{A}_7\right)+
(1-\kappa)\left(\mathcal{A}_8+\mathcal{A}_9+\mathcal{A}_{10}\right).
\end{eqnarray*}

\begin{figure}
  \begin{center}
        \includegraphics[width=30pc,height=20pc]{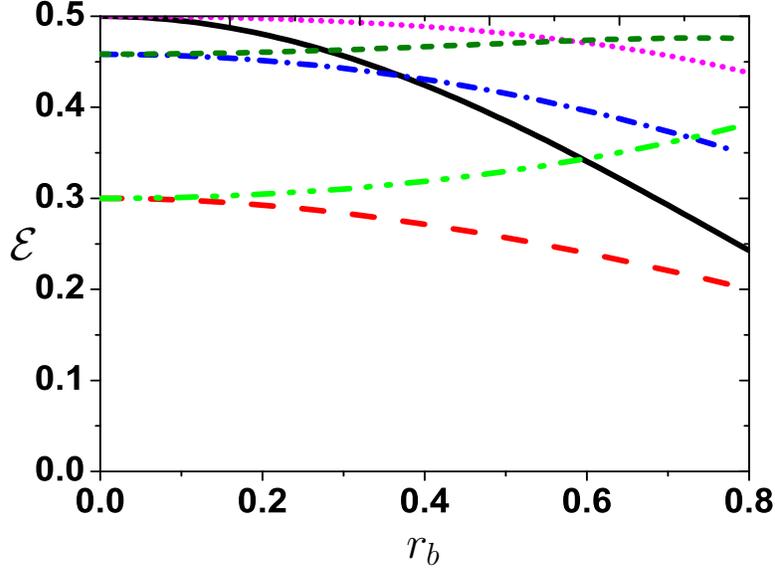}
     \put(-180,7){\Large$r_{b}$}
     \put(-330,120){\Large$\mathcal{E}$}
         \caption{ The  entanglement of the filtered accelerated state against the
         acceleration, where only the qubit is accelerated.
The dash, dash-dot,dot short dash and dash-dot-dot curves for
      $\kappa=0.1,0.3,0.5,0.7,0.9$, respectively, while the solid curve for the non-filtered state. }
  \end{center}
\end{figure}
Fig.(1), describes the behavior of entanglement of the accelerated
system between Alice and Bob after performing the filtering
process, where different values of the filtering strengths are
considered.  The effect of the filtering parameter $\kappa$,( only
Alice qubit is filtered), on the degree of entanglement is
investigated. It is clear that, for small values of $\kappa$, the
initial degree of entanglement ($r_{b}=0$) is smaller than that
depicted for the non-filtered case (solid-curve). However, as one
increases $\kappa\in[0,0.5]$, the decay rate of  entanglement
decreases and consequently the lower bounds of entanglement
increase. Moreover, for $\kappa=0.5$ the upper bound of
entanglement is larger than that displayed for the non-filtered
case. This behavior is changed dramatically for
$\kappa\in(0.5,1)$, where the initial  degree of entanglement  is
smaller than that shown for the non-filtered case. As $r_{b}$
increases, the entanglement increases and when the acceleration
goes to infinity, the entanglement is much better than the
non-filtered case.

\item{\it Bob's qutrit is filtered\\} In this case, the  final
state is given by

\begin{eqnarray}
\rho^{b-ac}_{t-F}&=&\frac{1}{N_{t-F}}\Bigl\{A_1\ket{00}\bra{00}+A_2\ket{01}\bra{01}+A_3\ket{02}\bra{02}
\nonumber\\
&&+A_4\ket{10}\bra{10}+A_5\ket{11}\bra{11}+A_6\ket{12}\bra{12})
\nonumber\\
&&+\sqrt{q}\left(A_7\ket{00}\bra{12}+A_8\ket{02}\bra{10}+A_9\ket{12}\bra{00}+A_{10}\ket{10}\bra{02}\right)\Bigr\}
\end{eqnarray}
where, $A_i,i=1..10$ are given by (7) and the normalization factor
$N_{t-F}=\sum_{i=1}^{6}A_i$

\begin{figure}
  \begin{center}
    \includegraphics[width=30pc,height=20pc]{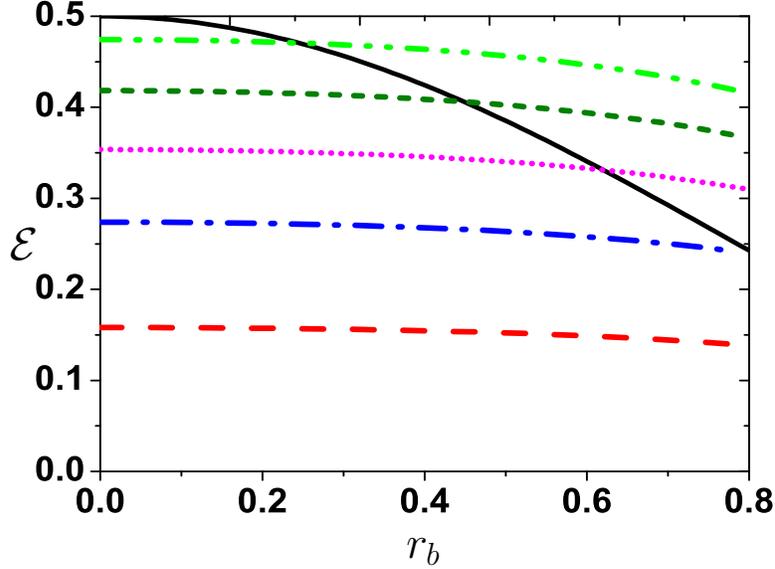}
      \put(-180,7){\Large $r_{b}$}
     \put(-330,120){\Large$\mathcal{E}$}\\
      \caption{ The sam as Fig.(1) but it is assumed that  the qutrit is filtered. The dash, dash-dot,
      dot short dash and dash-dot-dot curves for $\mathcal{Q}=0.1,0.3,0.5,0.7$ and $0.9$, respectively, and the sold curve for the n
      non-filtered case. }
  \end{center}
\end{figure}

Fig.(2) shows the behavior of entanglement when Bob's particle
(qutrit) is  filtered. It is clear that, for small values of the
filtering parameter $\mathcal{Q}$, the initial degree of
entanglement is smaller than that depicted for the non-filtered
case (solid curve). As one increases $\mathcal{Q}$, the
entanglement increases and slightly degreases at $r_{b}\to\infty$.
However  for $\mathcal{Q}\in(0.3,1)$, the upper bounds of
entanglement at $r_{b}\to\infty$ are always larger compared with
the non-filtered case. It is clear that, the entanglement can be
immunized from decaying by applying the filtering at the
decreasing points. For example, at $r_b\sim 0.6$, the entanglement
decay can be prevented, if the qutrit is filtered with a strength
$\mathcal{Q}=0.5$.

From  Figs.(1) and (2) one can concludes that, local filtering can
improve the degree of entanglement for the accelerating systems.
The long lived entanglement can be displayed if the qutrit is
filtered for any values of the strength's filter $\mathcal{Q}$.
For some particular values of $\kappa$ one can obtain a long-lived
entanglement. If the qubit is filtered, the entanglement increases
for $r_{b}\to\infty$, while it  slightly decreases if the qutrit
is filtered. Therefore, if the smaller dimensional subsystem is
accelerated and the larger dimensional subsystem is filtered, one
obtains a long-lived entanglement between the partners.  Moreover,
local filtering, not only can be used to immunize the entanglement
but also can be used to improve it. So, it can be considered as a
resource of  quantum purification to improve the efficiency of the
accelerated state.

\begin{figure}
  \begin{center}
        \includegraphics[width=20pc,height=15pc]{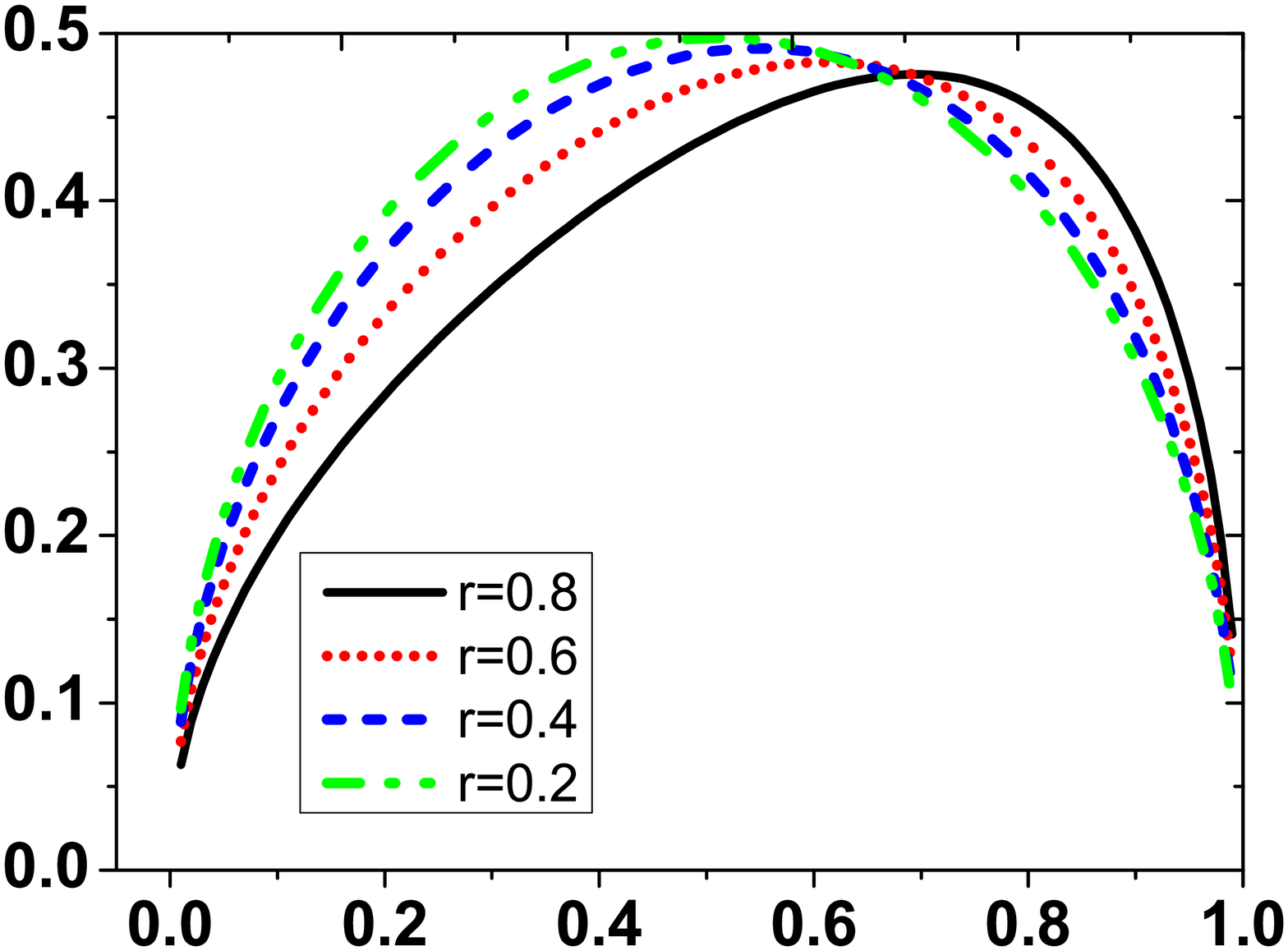}
     \put(-110,7){$k$}
     \put(-230,90){\Large$\mathcal{E}$}
     \put(-60,150){$(a)$}
 \includegraphics[width=20pc,height=15pc]{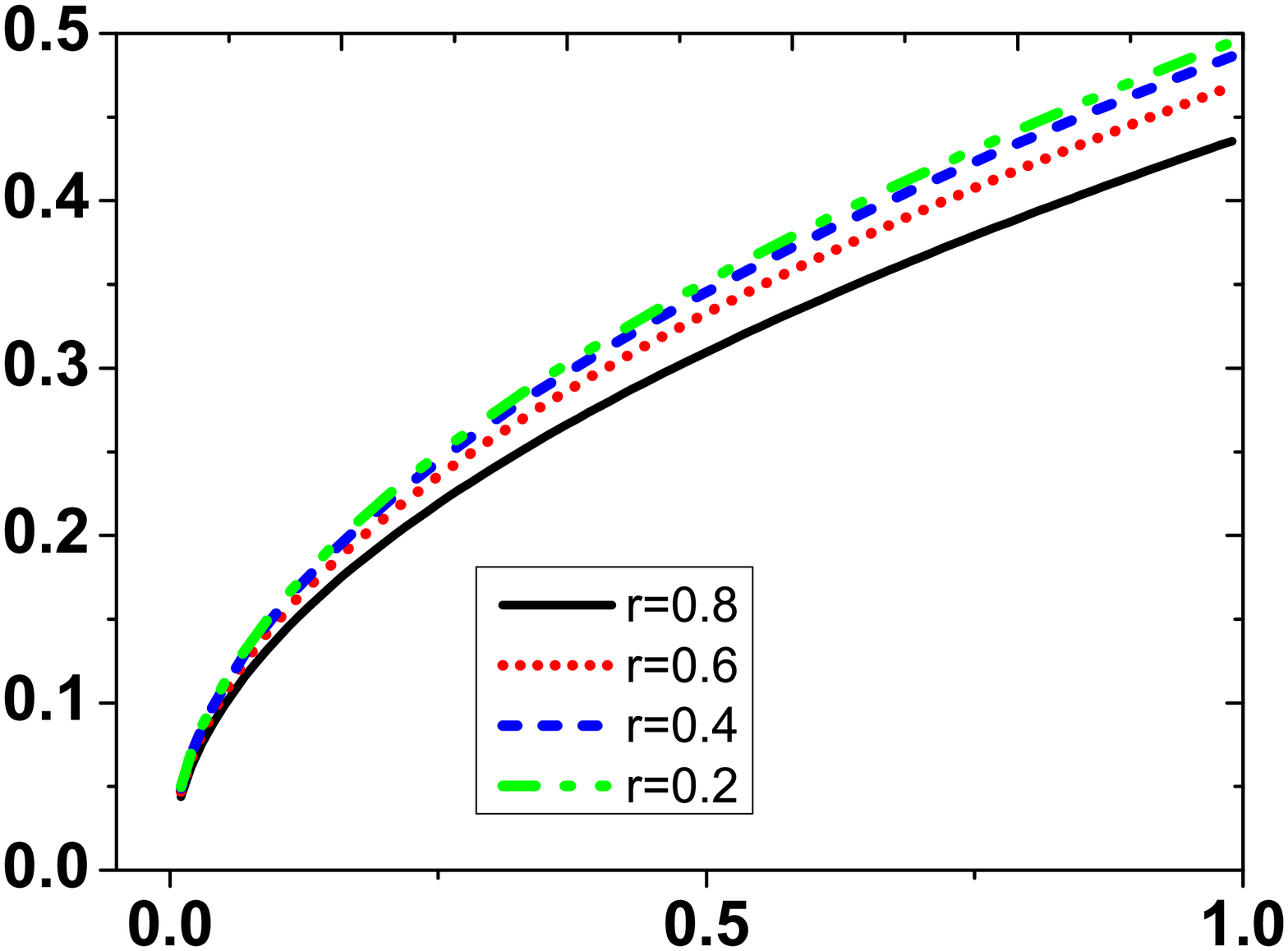}
         \put(-110,0){$\mathcal{Q}$}
     \put(-230,90){\Large$\mathcal{E}$}
     \put(-70,150){$(b)$}
      \caption{ The amount of entanglement   the strength of the
      filter $k$ for different values of the
      acceleration $r_b=r=0.8,0.6.0.4$ and $0.2$ . }
  \end{center}
\end{figure}
In Fig.(3), we investigate the behavior of entanglement against
the filtering strengths $\kappa$ and $\mathcal{Q}$, where
different values of $r_{b}$ are considered. In Fig.(3a), the
maximum values of entanglement depend on the initial acceleration
and the  strength of the filter. For  small accelerations  the
upper bounds of entanglement at a fixed values of
$\kappa\in(0,0.5)$ are much larger than those depicted for larger
acceleration. This behavior is changed for larger values of
$\kappa\in(0.5,1)$, where the larger acceleration, the smaller
upper bounds of entanglement. Fig.(3b), shows the effect of
qutrit's filter strength $\mathcal{Q}$ on the degree of
entanglement for different values of the accelerations. It is
clear that, the entanglement $\mathcal{E}$ increases as
$\mathcal{Q}$ increases and  its  upper bounds that  depend on the
initial acceleration, where they are small for larger
accelerations's values.

\begin{figure}
  \begin{center}
\includegraphics[width=30pc,height=20pc]{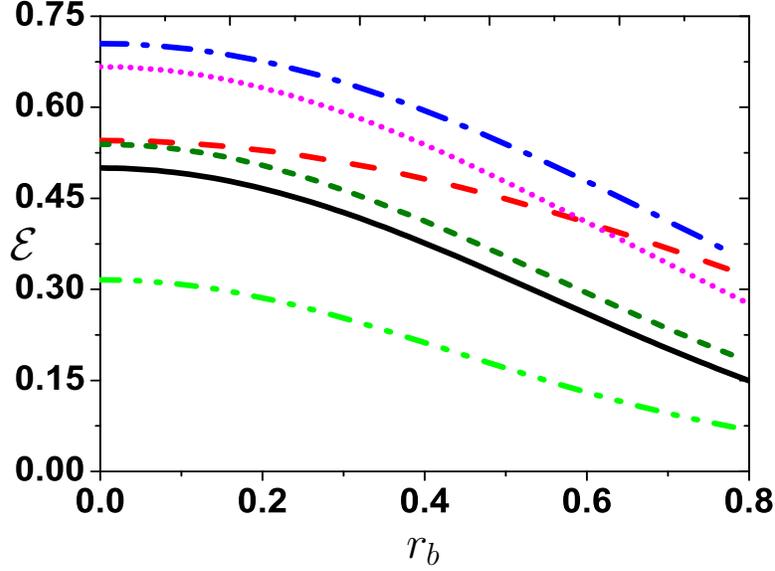}
     \put(-180,7){\Large$r_{b}$}
     \put(-330,120){\Large$\mathcal{E}$}
          \caption{ The same as Fig.(1) but only the qutrit   is accelerated.
       }
  \end{center}
\end{figure}

\subsection{Bob's qutrit is accelerated}
In this case, the final state between Alice and Bob is given by
Eq.(8). We consider the following possibilities:
 \item{\it Alice 's  qubit
is filtered\\} For this case Alice uses the filter defined by
Eq.(11) to filter her qubit. After finishing the filtering
process, the partners Alice and Bob share the following state.

\begin{eqnarray}
\rho^{t-ac}_{b-F}&=&\frac{1}{\tilde{N}_{b-F}}\Bigl\{\kappa
(\mathcal{B}_1\ket{00}\bra{00}+\mathcal{B}_2\ket{01}\bra{01}+\mathcal{B}_3\ket{02}\bra{02})
\nonumber\\
&&+(1-\kappa)(\mathcal{B}_4\ket{10}\bra{10}+
\mathcal{B}_5\ket{11}\bra{11}+\mathcal{B}_6\ket{12}\bra{12})
 \nonumber\\
&&+\sqrt{\kappa}\sqrt{1-\kappa}(\mathcal{B}_7\ket{02}\bra{10}+\mathcal{B}_8\ket{10}\bra{02}+
\mathcal{B}_9\ket{12}\bra{00}+\mathcal{B}_{10}\ket{00}\bra{12})\Bigr\}
\end{eqnarray}
where, $\mathcal{B}_i$ are given by (9) and the normalized factor
is defined as,
\begin{eqnarray*}
\tilde{N}_{b-F}=\kappa\sum_{i=1}^{3}\mathcal{B}_i+(1-\kappa)\sum_{i=4}^{6}\mathcal{B}_i
\end{eqnarray*}
Fig(4) describes the behavior of entanglement if the qutrit is
accelerated and the qubit is filtered. It is clear that, the
initial  degree of entanglement of the accelerated state is much
better than the non-filtering case for $0<\kappa<0.7$. The upper
bounds of $\mathcal{E}$ are lager for $\kappa< 0.5$. However for
farther values of $\kappa$ the upper bounds of entanglement
decrease. For larger values of $\kappa>0.8$, the degree of
entanglement is smaller than that displayed for the non-filtered
case (solid curves).

\item{\it Bob's  qutrit is filtered\\} In this case we assume that
only Bob particle (qutrit) is accelerated and has the ability to
filter it. The final state of these two operations is given by,

\begin{eqnarray}
\rho^{q_t-ac}_{t-F}&=&\frac{1}{\tilde{N}_{t-F}}\Bigl\{B_1\ket{00}\bra{00}+B_2\ket{01}\bra{01}+B_3\ket{02}\bra{02}+B_4\ket{10}\bra{10}+
B_5\ket{11}\bra{11} \nonumber\\
&&+\sqrt{q}(B_7\ket{02}\bra{10}+B_8\ket{10}\bra{02}+B_9\ket{12}\bra{00}+B_{10}\ket{00}\bra{12})\Bigr\}
\end{eqnarray}
where $B_i, i=...10$ are given by (6) and
$\tilde{N}_{t-F}=\sum_{i=1}^{5}\mathcal{B}_i$

\begin{figure}
  \begin{center}
     \includegraphics[width=30pc,height=20pc]{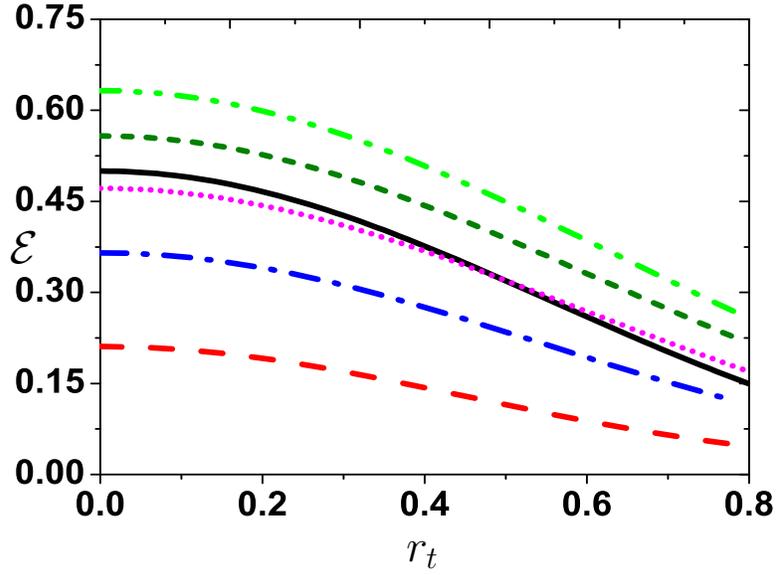}
     \put(-180,7){\Large$r_{t}$}
     \put(-330,120){\Large$\mathcal{E}$}
      \caption{ The same as Fig.(2) but it is assumed only  the qutrit   is accelerated.
       }
  \end{center}
\end{figure}

In Fig.(5),  the effect of the filtered qutrit on the entanglement
of the accelerated state is investigated. It is clear that, the
entanglement increases as the filtering strength $\mathcal{Q}$
increases. For small values of $\mathcal{Q}\in(0,~0.5)$, the
filter process can not improve $\mathcal{E}$. However, for small
range of $r_{t}\in(0.45,8)$ the entanglement can be improved if we
set $\mathcal{Q}=0.5$. Moreover for any value of
$\mathcal{Q}\in(0.5,1)$ the degree of entanglement is much better
than that shown for the  non filtered case (solid curve).

\begin{figure}
  \begin{center}
        \includegraphics[width=20pc,height=15pc]{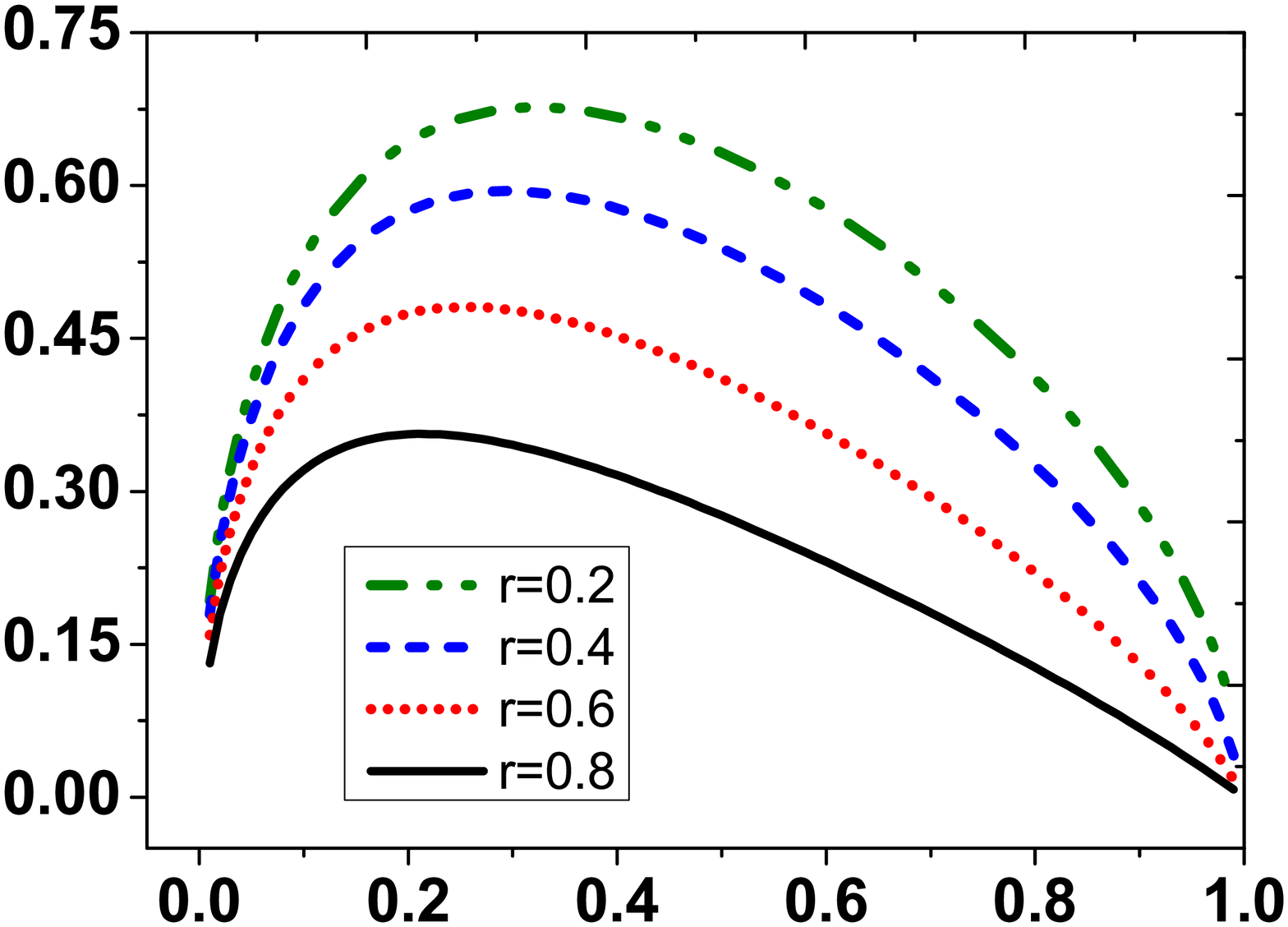}
     \put(-110,7){$k$}
     \put(-230,90){\Large$\mathcal{E}$}
      \put(-60,150){$(a)$}
     \includegraphics[width=20pc,height=15pc]{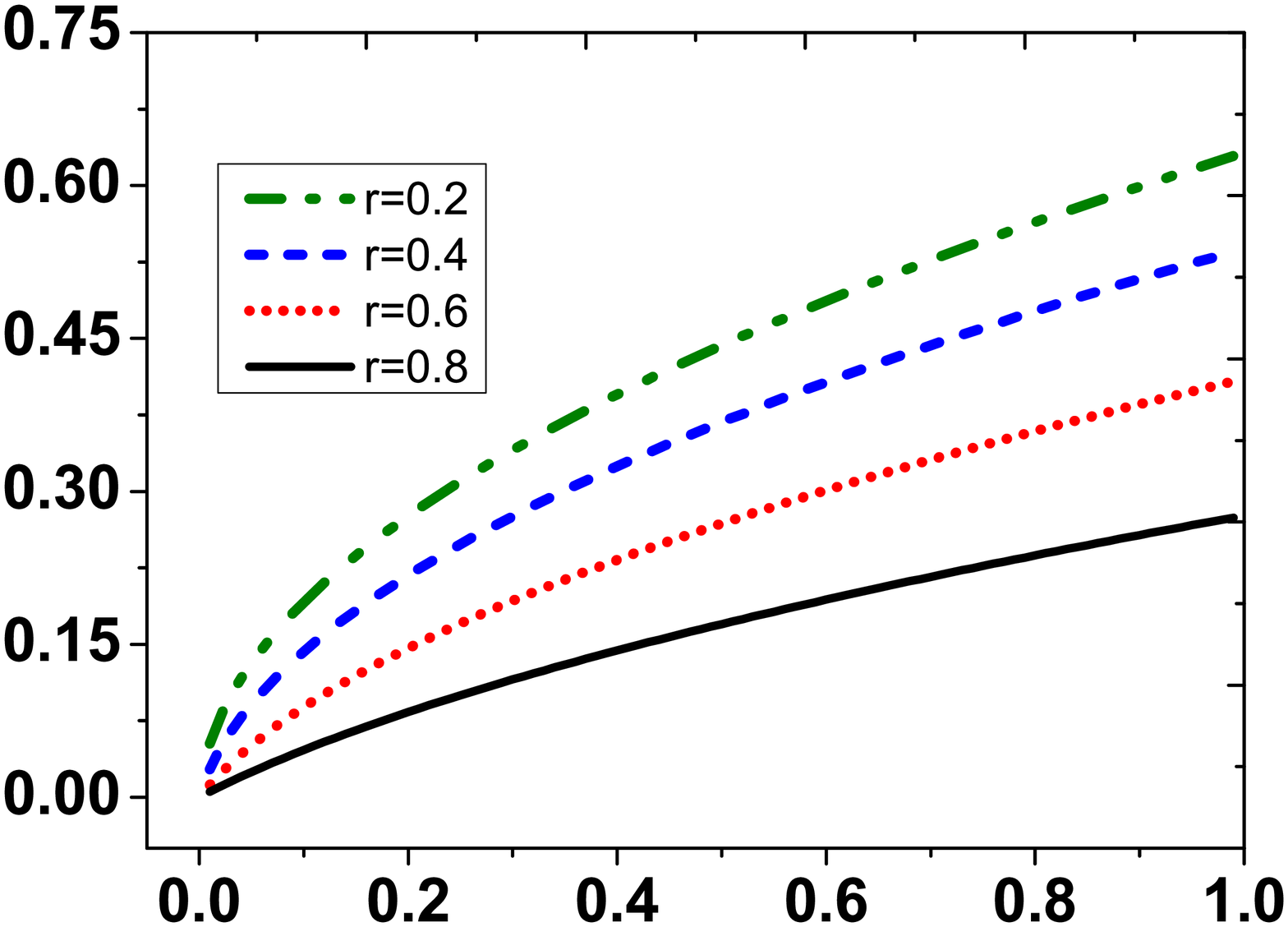}
    \put(-110,7){$\mathcal{Q}$}
     \put(-230,90){\Large$\mathcal{E}$}
      \put(-60,150){$(b)$}
      \caption{ The survival amount of entanglement for one parameter family against
      Filtering parameter (a) qubit filter strength, $\kappa$  and (b) qutrit filter strength $\mathcal{Q}$.  }
  \end{center}
\end{figure}
Fig.(6) describes the behavior of entanglement against the
filter's strengths, where we consider some fixed values of the
accelerations $r_t$. If only the qubit is filtered, one can see
that the entanglement increases as $\kappa$ increases in intervals
depending on the initial value of the accelerations. The maximum
values of entanglement are large for small values of the
acceleration. However, for further values of $\kappa$, the
entanglement decreases. These phenomena can be seen in Fig.(6a).
Meanwhile, if the qutrit is filtered, the entanglement increases
as $\mathcal{Q}$ increases. Also, the maximum values of
$\mathcal{E}$ are reached for small values of $r_{t}$ as shown in
Fig.(6b).

From Figs.(1) and (5), one can conclude that: it is possible to
improve the entanglement between the partners if one user has
accelerated his(her) subsystem. The improvement of the degree of
entanglement depends on  the initial acceleration, where for small
accelerations one can improve it by controlling the filtering
strength.  For fixed accelerations, the maximum values of the
entanglement can be reached as one increases the qutrit's filter
strength. The longed lived entanglement can be obtained if the
smaller dimension subsystem (qubit) is accelerated and the larger
dimensional subsystem (qutrit) is filtered. If the qutrit is
accelerated one can increase the initial entanglement by filtering
either the qubit or the qutrit. Meanwhile, the initial
entanglement can not be increased if the small dimension subsystem
(qubit )is accelerated and either the qubit or the qutrit is
filtered.

Comparing Figs.(3) and (6), we can see that for any subsystem
accelerated  the entanglement can be increased as the qutrit's
filter strength is increased. On the other hand, if any subsystem
is accelerated and  Alice has the ability to filter her qubit,
then the maximum values of entanglement  depend on the initial
acceleration and the values of the qubit's filter strength

\end{itemize}

\section{Conclusion}

In this contribution,  the possibility of improving the
entanglement of an accelerated  state consists of two different
dimensional subsystems is investigated. A state  composed of qubit
(two-dimensional) and qutrit (three dimensional) is considered to
illustrate this idea . This state is described by only one
parameter and  it is known  one-parameter family. Different cases
are considered, where it is assumed that only one particle is
accelerated. The final state in Minkowski space is obtained
analytically for all cases.  To improve the entanglement or
decreasing the rate of entanglement decay, we consider only one
filtered subsystem  (qubit/qutrit).

In  this context, it assumed that only one of the subsystem is
accelerated, while the filtering process is performed on the qubit
(small dimension subsystem)  or the qutrit (large dimension
subsystem), the qubit. It is shown that, local filtering can
increase the upper bounds of entanglement of the accelerated
system. The maximum bounds of entanglement depend on the values of
the acceleration, filters' strengths  and the dimensions of the
subsystem which is accelerated/filtered.

The obtained results show that, for a fixed acceleration of the
qubit (small dimension subsystem), the entanglement of the
filtered state increases as the qutrit's filter strength
increases, whilst if the qubit is filtered, then the entanglement
increases  over an interval of the strength depending on the
initial acceleration. However, for larger values of the qubit's
filter strength, the entanglement of the  large initial
accelerations increases. On the other hand, if the qutrit is
accelerated, then the increasing rate of entanglement  at  fixed
values of accelerations is much larger than that depicted for the
previous case (the qubit is accelerated). The larger values of
entanglement's bounds are drown for larger values of the qutrit's
 filter strength, while they are seen for smaller values of the
qubit's filter strength

If the smallest dimension subsystem  is  accelerated, then the
entanglement can not increase its initial values. One the other
hand, for smaller acceleration, the  entanglement bounds of the
filtered state  are much larger than those compared  with a large
acceleration. If the smaller dimensional subsystem is accelerated
and filtered, the maximum entanglement of the filtered state
depends on the initial acceleration and the  value of the filter
strength. Meanwhile, for accelerated qubit, one can obtain a long
lived entanglement between the accelerated partners, where its
upper bounds always increase as the filter strength of the qutrit
 increases.

{\it In conclusion:} local filtering can  be considered as a
resource of  purifying the accelerated states, where it is
possible to retrieve the lost entanglement. Local filtering play
the same role played by the weak measurements to improve the
entanglement. For noise environment, local filtering has the
ability to retrieve the lost entanglement parabolisticaly, while
for accelerated system in addition to retrieve the lost
entanglement, it can increase it.  If the smallest dimension
subsystem is accelerated, one can not only recoup the lost
entanglement but also a long-lived entanglement can be generated
by filtering the larger dimension subsystem. However, if the
largest subsystem is accelerated, one can in addition to
retrieving the lost entanglement, increase the upper bounds of
entanglement  by filtering any subsystem (qubit/qutrit).

\end{document}